\documentstyle[12pt,aasms4]{article}

\begin{document}
\title{Far-UV Observations of the Galactic Supersoft Binary
RX~J0019.8+2156 (QR And) \footnote{Based on observations made with the
NASA-CNES-CSA Far Ultraviolet Spectroscopic Explorer. FUSE is operated for
NASA by the Johns Hopkins University under NASA contract NAS5-3298}} 
\author{J.B. Hutchings\altaffilmark{2},
D. Crampton\altaffilmark{2},
A.P. Cowley\altaffilmark{3}, 
P.C. Schmidtke\altaffilmark{3}, and
A.W. Fullerton\altaffilmark{4,}\altaffilmark{5} }

\altaffiltext{2}{Herzberg Institute of Astrophysics, National Research
Council of Canada,\\ Victoria, B.C. V9E 2E7, Canada; email:
john.hutchings@nrc.ca, david.crampton@nrc.ca } 

\altaffiltext{3}{Department of Physics \& Astronomy, Arizona State
University, Tempe, AZ 85287-1504; email: anne.cowley@asu.edu,
paul.schmidtke@asu.edu } 
 
\altaffiltext{4}{Dept of Physics \& Astronomy, University of Victoria,
P.O. Box 3055, Victoria, B.C., V8W 3P6, Canada} 

\altaffiltext{5}{Dept of Physics \& Astronomy, Johns Hopkins University,
3400 N.\ Charles St., Baltimore, MD 21286; email: awf@pha.jhu.edu} 

\begin{abstract}

FUSE spectra were obtained of the supersoft X-ray binary RX~J0019.8+2156
(QR And) during 16 consecutive spacecraft orbits, covering the binary
orbit (P=15.85 hr) with about 0.2 phase overlap.  The spectrum is
dominated by strong H$_2$ absorption (column density $\sim$10$^{20}$ g
cm$^{-2}$), which appears at the velocity different from other
interstellar absorption lines and may be partially circumbinary. This
absorption makes study of spectral features from the binary system
difficult.  The only well-detected emission lines are He~II, 1085\AA, and
O~VI, 1032\AA, (the other line of the O~VI doublet, 1037\AA, is largely
obscured by strong H$_2$ absorption).  The O~VI shows a P Cygni profile
that varies in velocity and strength with binary phase.  We compare this
with similar changes seen in Balmer line profiles.  We extract the FUV
light curve and compare it with the optical light curve.  There is an
eclipse in both wavelength regions, but the FUV minimum lasts much longer,
well beyond the visible light egress.  The FUV results are discussed in
connection with the binary model and mass flows within the system. 

\end{abstract}


\section{Introduction}

RX~J0019.8+2156 (hereafter called by its variable star designation, QR
And) is one of the few identified Galactic supersoft X-ray binaries,
although numerous examples have been found in the Magellanic Clouds and
M31 (see Greiner 1996).  The system was first noted in the $ROSAT$ All-Sky
Survey and described by Beuermann et al.\ (1995).  They found this 12th
magnitude system to show $\sim$0.5 mag eclipses and radial velocity
variations which fit an orbital period of 15.85 hrs.  Archival
photographic plates show the orbital period has been stable for over 40
years (Greiner \& Wenzel 1995), although there are irregular changes of a
few tenths of a magnitude in the overall brightness level of the system
(e.g. Cowley et al. 1998, Deufel et al. 1999) 

The system is at Galactic latitude $-40^{\circ}$, and hence shows little
evidence of interstellar reddening, with mean $U-B\sim-0.7$ (Beuermann et
al. 1995) and quite weak interstellar lines of Ca~II and Na~I (McGrath et
al. 2001).  This, with the mean absolute magnitude of supersoft binary
systems, places its distance at $\sim$2 kpc, so it lies well out of the
Galactic plane. 

Optical spectra show very strong emission lines of He~II (virtually
all of the He~II lines in the Multiplet Table are identified).  Balmer
absorption lines are present through about half of the orbital cycle and
indicate a strong outflow of material with velocities of $\sim-200$ to
$-600$ km s$^{-1}$.  Like several other supersoft X-ray binaries, QR And
also shows bi-polar emission-line jets (velocity $\sim\pm800$ km s$^{-1}$)
in He~II (4686\AA) and the lower Balmer lines (e.g. Becker et al. 1998,
Cowley et al. 1998, Deufel et al. 1999).  These jet features come and go
on timescales of weeks to months.  Since the Balmer outflow is
phase-dependent (and persistent?), this is likely a constant disk flow
that is not strongly related to the jet activity. 
 
There has been some disagreement about the appropriate model for QR And,
with various authors deriving quite different values for the orbital
inclination, and hence the stellar masses.  Beuermann et al.\ (1995)
concluded that the inclination must be quite low ($i\sim20^{\circ}$),
based partially on the assumptions that the masses be similar and
consistent with a white dwarf.  On the other hand, Meyer-Hofmeister et
al.\ (1998) analyzed the light curve and presented models which indicate
an inclination near $i=55^{\circ}$, while Tomov et al.\ (1998) suggested
an even larger value ($i\sim79^{\circ}$) also based on the light curve. 
The adopted inclination is important since it ultimately confines the
range of possible masses, and hence the model and evolution of the system.
>From an analysis of the emission-line velocities and those of the
bi-polar jets, Becker et al.\ (1998) found that the jet lines show the
same velocity amplitude and phasing as He~II, 4686\AA, and hence concluded
that the He~II lines must reveal the motion of the compact star where the
jets originate.  Further, from analysis of the outflow line profiles they
found that the orbital inclination must lie in the range
$35^{\circ}<i<60^{\circ}$.  For the purpose of our discussion below, we
have adopted these conclusions of Becker et al.\ and assume a model 
similar to that discussed by Meyer-Hofmeister et al.

O~VI emission (3811, 3835, 5290\AA) is found in all supersoft X-ray
binaries and is potentially interesting as its high ionization may
indicate an origin close to the accretion disk center.  However, study of
these lines is compromised by blends with He~II at 3813\AA\ and 3834\AA,
~and with [Fe~XIV] at 5303\AA.  This suggested that it would be
particularly interesting to observe the supersoft systems in the far
ultraviolet with FUSE since the O~VI resonance lines (1032, 1037\AA) lie
within its wavelength range. 

\section{FUSE Observations} 

An overall description of the FUSE data is given by Sahnow et al.\ (2000).
Our FUSE observations were carried out during 16 successive spacecraft
orbits on 2000 July 27-28.  The large apertures
(30$^{\prime\prime}\times30^{\prime\prime}$) were used.  No unusual
anomalies were apparent during the observation sequence.  The observations
were taken in time-tag mode, and the 16 separate spectra cover slightly
more than one binary orbit.  Table 1 lists the details of the
observations. 

The spectra were processed by the CALFUSE pipeline and retrieved from the
archive.  Later, the data were reprocessed using an updated version
(1.8.7)  of the pipeline processes.  The results were the same to within
about 1\%. 

Surprisingly, the spectra were found to be heavily absorbed by molecular
hydrogen (H$_2$).  The amount of absorption is characteristic of more
reddened targets (E$_{B-V}=$0.5 or more), whereas, as noted above, QR And
has quite blue colors.  The spectra also contain airglow emissions (see
Feldman et al.\ 2001) which are much stronger on the daylight side of the
FUSE orbit.  Spectra 6, 8, 10, and 12 were taken almost entirely during
orbital night, and thus they provide a way to assess the airglow
contamination of the other spectra.  Fortunately, the spectral region near
the O~VI 1032\AA\ emission line is free of airglow lines, so that our
analysis of this region could be performed on data from both orbital night
and day without editing. 

\section{Spectral Energy Distribution and H$_2$ Absorption} 

In order to assess the contamination due to H$_2$ absorption, we used
simple absorption models with a range of column densities.  We also
derived an empirical absorption model by using a median spectrum of the
region of interest around the O~VI resonance lines.  The 1032\AA\ line is
free of strong H$_2$ absorptions, but the 1037\AA\ line is almost
completely overlaid by saturated absorption. 

Figure 1 shows the average spectrum with different amounts of smoothing
and airglow removal, as well as the H$_2$ model from our grid that fits it
most closely.  The H$_2$ absorption strength in the FUSE wavelength range
is loosely correlated with E$_{B-V}$ and was compared with several B0
stars with E$_{B-V}$ ranging from 0.15 to 0.5 magnitudes which had been
observed with FUSE.  The values for QR And correspond to values of
E$_{B-V}\sim$0.5.  However, Beuermann et al.\ (1995) estimated a value of
E$_{B-V}\sim0.10$ for QR And, based on the strength of its 2200\AA\
absorption feature.  The colors of QR And ($U-B=-0.7$ and $B-V=0.01$) also
show the system to be very blue, so the E$_{B-V}$ cannot be 0.5 and the
H$_2$ absorption strength is unusually high for the reddening.  Since the
system is losing matter via disk and jet outflow, it is worth considering
whether some of the H$_2$ absorption is due to material surrounding the
binary.  We estimate the column density causing the H$_2$ absorption to be
$\sim$10$^{20}$ cm$^{-2}$.  Beuermann et al.\ (1995) estimated an
absorbing column of $4\times10^{20}$ H-atoms cm$^{-2}$ from $ROSAT$ data
for a combination of Galactic foreground absorption plus any possible
intrinsic absorption. 

To determine if the H$_2$ absorptions are interstellar or circumstellar,
we compared their velocities with Ca~II and Na~I interstellar lines and
with QR And's systemic velocity, both derived from optical spectra.  From
the FUSE spectra we measure the heliocentric velocity of H$_2$ features
near the O~VI lines to be $-16\pm2.5$ km s$^{-1}$, with no variation
from spectrum to spectrum.  The absolute wavelength precision depends on
placing the star accurately in the (guided) LiF1 channel aperture.  A
precision of 1 arcsec is expected, which corresponds to $\pm$3.4 km s$^{-1}$. 
To further check the FUSE wavelength zero-point (possibly due to grating
movement uncertainty) we measured the Ly$\beta$ airglow wavelength. This 
tracks the orbital velocity of FUSE along the boresight to within 2 km 
s$^{-1}$, so we conclude that velocities are correct as measured. 

An MMT spectrum obtained 1995 Oct.\ 12 shows the mean velocity from
interstellar absorption lines of Ca~II and Na~I to be $+4\pm4$ km
s$^{-1}$.  McGrath et al.\ (2001) found the systemic velocity of QR And to
be $\sim-41\pm3$ km s$^{-1}$ (Beuermann et al.\ (1995) gave $-59\pm2$ km
s$^{-1}$ from lower resolution data).  The measured H$_2$ velocity of
$-16$ km s$^{-1}$ lies between the ISM and binary velocities. In view of
the unusual H$_2$ strength, it thus seems likely that it is partly
associated with the binary system - perhaps formed during a PN episode and
well-removed spatially from the current ionizing radiation from supersoft
X-ray activity. 

\section{Far Ultraviolet and Optical Light Curves}

The FUSE data were used to construct the FUV light curves shown in Figure
2.  FUSE consists of four separate telescopes that are kept co-aligned by
compensating for modelled mechanical distortions around the orbit (e.g.
Moos et al.\ 2000).  Only one telescope (LiF-1) is used for tracking.  The
extraction of signal is thus uncertain in the other three telescopes if
the signal is weak and the position of the target in the
(30$^{\prime\prime}$) aperture drifts by a great deal.  The LiF1A (longer
wavelengths) are also subject to a detector wire shadow (the `worm') that
may affect photometric accuracy.  However, the LiF2A (longer wavelength:
1090--1180\AA) channel has little drift and the highest signal level.
Thus, we used the fluxes from the LiF1A (990--1080\AA) channel and the
LiF2A  channel to derive two independent light curves.  The continuum
fluxes were estimated by summing these spectral regions after removal of
airglow lines and interpolating across the deep H$_2$ absorptions.  These
are shown as a function of phase in Fig.\ 2, displayed as magnitudes.
There is good correspondence between the two ultraviolet regions.  The
count rates for all channels are steady and show no variations that might
indicate the star was drifting out of the aperture.  The spectral images
show that the extraction windows appear to cover the data well.  In the
binary phases where there are overlapping FUSE observations (phases
0.5--0.7), the fluxes are in good agreement. 

The FUSE fluxes were rechecked by one of us (AWF) on the raw data using
the current version of CALFUSE (1.8.7).  The calibrated spectra include
the total counts associated with each pixel (i.e., the sum across the
width of the spectrum in the spatial direction).  A light curve from the
total signal over the interval 1102--1133\AA\ was derived from the LiF1B
channel.  This region does not have any strong airglow lines and avoids
the LiF1B channel `worm'.  The result is essentially the same as the
1090--1180\AA\ LiF2A curve shown in Fig.\ 2.  To check CALFUSE further,
spectra for 5-minute intervals were extracted from the raw data, and count
rates were determined over the same 1102--1133\AA\ wavelength region.
These light curves are also very similar to the FUV light curves shown in
Fig.\ 2. 

Comparing the two FUV light curves in Fig.\ 2, we find that overall
amplitude is larger by $\sim0.03$ magnitudes in the longer wavelength
channel.  This is likely to be caused by H$_2$ line blanketing that
lowered the continuum in the shorter wavelength region where the
absorptions are more crowded (see Fig.\ 1), rather than being a real color
change.  We note that the visible light shows almost no color changes
through eclipse (e.g. Cowley et al. 1998, Deufel et al. 1999). 

For comparison of the optical and far-ultraviolet fluxes, Fig.\ 2 shows
two $V$-band light curves together with the two derived from the FUSE
data.  The nearly complete optical light curve (small filled circles) was
obtained in 1995 September and is fully described by McGrath et al.\
(2001).  It is plotted with 9-point smoothing applied.  In the unbinned
optical data, flickering is present throughout the orbital cycle, and
quasi-periodic variations (P $\sim$1.8 hours) are prominent at orbital
phases 0.2 to 0.6.  The second optical light curve (small open circles)
shows $V$ data taken on 2000 July 28 by one of us (PCS) at Braeside
Observatory in Flagstaff, Arizona, with monitoring beginning just hours
after completion of the FUSE observations.  The orbital phase coverage is
rather limited, but it includes most of primary eclipse.  The short-term
fluctuations are of much smaller amplitude than in the 1995 data.  In
2000, QR And was $\sim$0.1 magnitude brighter than in the 1995 data, but
variations in the mean brightness level are well-known in this system (cf.
Will \& Barwig 1996, Matsumoto 1996, Cowley et al. 1998, Deufel et al.
1999).  The optical data shown here illustrate the typical behavior of
this source. 

There are very obvious differences between the optical and FUV light
curves in the phases following minimum light.  While the ingress is
similar in both wavelength regions, the FUV light curve shows a minimum
that lasts from phase 0.95 through 0.25, much longer than the minimum
observed in optical light.  Thus, we appear to have an eclipse of the
hottest regions of the disk that extends well beyond that of the cooler
regions seen in optical wavelengths.  If this extended minimum is real, it
could indicate important disk structures in the system which presumably
effect the FUV radiation.  However, we would really like to observe
another minimum to be certain there was no FUSE instrumental problem.  The
`secondary eclipse' dip present in the visible light curve is not seen in
the FUV, but a small dip does occur at phase $\sim$0.6. 

If the FUV extended eclipse represents a permanent feature, we might
expect to find a gradual change in the light curve  as one observes at
progressively shorter wavelengths.  The best we can do at the present time
is to examine other light curves available in the literature.  Beuermann
et al.\ (1995) show light curves at 1400--1500\AA\ and X-ray data from
$ROSAT$-PSPC.  Replotting these data with a more extended vertical scale
we find both the duration of minimum and the phases of ingress/egress to
be consistent with our optical light curve and that of Will \& Barwig
(1996).  Similarly the $IUE$ data shown by Gaensicke, Beuermann, \& de
Martino (1996) for 3000\AA\ and 1250\AA\ when replotted also show similar
phases of ingress/egress and duration of minimum flux.  We were unable to
find any indication from published ultraviolet data of other
extended-duration eclipses such as seen in the FUSE data.  However,
Gaensicke et al.\ (2000) show $ROSAT$-HRI X-ray data obtained between 1997
Dec.\ 2 and 1998 Jan.\ 10 in which there is no evidence of any orbital
variation.  Thus, changes do occur in QR And which severely modify the
light curve.  Furthermore, Meyer-Hofmeister et al.\ (1998) found that
important changes in the disk-rim structures can occur on timescales as
short as the orbital period.  However, without additional FUSE data, we
don't know if the extended minimum we observed in the FUV light curve was
a transient event or due to a more permanent structure in the binary
system. 

\section{FUV Spectrum}

The spectra were examined carefully for the presence of high ionization
lines in the FUSE range.  While the H$_2$ contamination makes this
difficult, we find no evidence of N~III, C~III, S~IV, Si~V, or P~V.
However, O~VI is clearly present and has a complex and variable line
profile that we discuss in detail below.  The only other line we detect is
He~II 1085\AA.  This line is close to two airglow lines of N~II, but both
the median of all spectra and the sum of the night spectra show there is a
broad He~II emission line present in addition to the narrow and variable
airglow lines.  Due to the airglow contamination and the weakness of
He~II, we are unable to study its behavior as a function of binary orbital
phase.  Nevertheless, it is clear that the line does not have the P Cygni
absorption profile found in the O~VI lines.  This is consistent with the
purely emission profiles of the visible Pickering He~II lines reported by
McGrath et al.\ (2001).  We note that in ORFEUS~II observations
(900--1200\AA) no stellar emission lines were detected (Gaensicke et al.
2000). 

In the optical region, lines of C~III and C~IV (and possibly N~III and
N~V) are weakly present (McGrath et al. 2001), but IUE spectra of QR And
show no evidence of the resonance lines of C~IV at 1550\AA\ (Beuermann et
al. 1995).  The only emission feature in the IUE data is He~II 1640\AA. 
In the FUSE spectra, C~III at 1175\AA\ is not detected, even though it is
clear of H$_2$ and airglow.  We set an upper limit on its equivalent width
of less than 0.1\AA.  It appears that carbon, and possibly nitrogen, have
low abundances in QR And. 

\subsection{O~VI}

The strongest emission line in the FUSE range is O~VI, 1032\AA.  This
whole feature is fortunately free of airglow or major H$_2$ absorption
(there are some weak narrow absorptions but they are comparable with the
noise levels in individual spectra and very much narrower and weaker than
the stellar lines).  The line has a P Cygni profile which changes around
the binary orbit, as shown in Figure 3.  By subtracting an H$_2$ model and
looking at differences from the median of all spectra, we can also see the
edges of the absorption and emission of the O~VI 1037\AA\ line, which
gives us added confidence that the changes we see in the 1032\AA\ line are
real. 

The extensive spectral coverage of the optical region by McGrath et al.\
(2001) shows that the lines near Balmer wavelengths have P Cygni profiles
that change markedly around the orbit.  (Note that the emission portion of
these lines is a blend of H+He~II, as one can tell by comparing their
strengths with the non-blended, adjacent He~II Pickering emission lines.
However, the variable-strength absorption component is due only to H.)
Fig.\ 3 shows the changes in both the FUSE O~VI line and the optical
`Balmer' lines during similar phases (computed using the optical ephemeris
of McGrath et al.)  The FUSE data are smoothed by 91 points, which makes
the resolution comparable with the optical, without losing the significant
changes in the broad line profiles.  In order to compare the O~VI and
`Balmer' lines, the optical spectra were averaged from numerous spectra
within the phase bins specified, but were not smoothed.  The profiles in
Fig.\ 3 were normalized to the local continuum, and measurements were made
of the centroids and fluxes of the emission and absorption components as
shown in Figure 4.  The values plotted in Figure 4 have measured total
scatter of $\pm$20\% for the EW values, and velocities are reproducible
to $\sim$50 km s$^{-1}$, taking into account uncertainties in continuum
fitting.

The absorption outflow velocities are similar for all three lines shown in
Fig.\ 4, but the variations in absorption strength differ.  O~VI has
absorption at all phases, while little or no Balmer absorption is present
from phases 0.7 to 0.0.  The O~VI and H absorption strengths thus show no
binary phase correlation in Fig.\ 4, while the O~VI emission and
absorption do follow the phase variations of the Balmer lines, albeit with
a smaller amplitude.  This suggests that there may be significant
azimuthal variations of the ionization of the outflowing material.  In
this connection we note that FUSE data show anomalous O~VI presence in the
outer regions of OB star stellar winds (Crowther, Bianchi: in preparation)
whose ionization may be connected with X-ray flux.  Further work is needed
to understand these phenomena properly. 

If the velocity of the O~VI emission is due to orbital motion of the
compact star, maximum velocity should occur at photometric phase 0.25.
However, the maximum occurs $\sim$0.1P earlier, almost certainly due to
the very strong P Cyg absorption present between phases 0.0 and 0.2.  The
O~VI velocities are very similar to those found by McGrath et al.\ for the
blended H+He~II emission lines.  These authors suggest that only the pure
He~II lines (i.e. those with no trace of overlying H absorption) reveal
the orbital motion of the compact star, the same conclusion as was reached
by Becker et al.\ (1998).  The variable strength of the hydrogen
absorption modifies the emission-line profiles, hence distorting their
velocity curves. 

Changes in the line profiles are shown in Figure 5.  Differences from the
phase bin showing emission-only `Balmer' lines are plotted, so the figure
shows the changes in the absorption with phase.  The O~VI absorption is
more complex and is seen at all binary phases.  It is weakest during the
low parts of the FUV light curve, suggesting that the O~VI absorption is
associated with the FUV continuum source. 

\section{Discussion and Summary}

The optical spectra of McGrath et al.\ show that there are significant
outflows from the system, with strong azimuthal variations.  Generally
speaking, the outflows have velocities of several hundred km s$^{-1}$ and
are seen most strongly through mid-eclipse and egress, but not seen during
eclipse ingress.  In the FUSE data, the O~VI resonance lines show an
outflow of very hot gas at all phases, but with velocities similar,
although somewhat lower, to the Balmer absorption. 

The emission and absorption features in O~VI and H+He~II show similar
velocity changes with binary phase.  These are not the same as the pure
He~II emission velocity curves which appear to trace the binary orbital
motion (e.g. Becker et al. 1998; McGrath et al. 2001).  Thus, the wind
from the binary has azimuthal velocity variations.  The relative
absorption strength also changes with binary phase, which may indicate
azimuthal ionization changes.  The production of O~VI in OB stars is
anomalous and may have some connection with X-ray ionization which could
also be occurring in QR And. 

We find that the high H$_2$ absorption column, which is anomalous for the
unreddened system color, has a different velocity from the atomic
interstellar lines, lying between the ISM and assumed binary systemic
velocity.  We suggest that it may partially arise in a circumbinary
location, possibly formed during the presumed PN episode that occurred
prior to the present evolutionary stage.  We note that circumbinary H$_2$
is detected around some planetary nebulae and that other degenerate
binaries we are studying with FUSE do not show appreciable H$_2$
absorption. 

Finally, the very odd FUV light curve shows an extended eclipse minimum
lasting about one third of the orbit, although only one eclipse was
observed by FUSE.  While it may be tempting to connect the extended FUV
eclipse with variable H$_2$ absorption, the FUV light curve was derived
from the longer FUSE wavelengths which are free of H$_2$ absorption. 
Furthermore, we have determined that there is no change in the column
density of H$_2$ or its velocity during these phases.  The difference
between the visible and FUV eclipse might be explained by a UV-bright spot
on the rim of the disk, where a mass-transfer stream impacts it (e.g.
models by Meyer-Hofmeister et al. 1998).  The off-axis location of such a
spot could give rise to an extended eclipse in the FUV.  Such a model
would imply that the effect becomes more visible as the wavelength of
observation decreases from visible to FUV.  Our examination of archival
photometry at UV and X-ray wavelengths suggests that the light curve may
show large long-term changes. 

If the O~VI line profile (both emission and absorption) arises in part (or
entirely) at this UV-bright location, the observed phasing of the O~VI
radial velocities is qualitatively as expected.  One can also imagine
different outflow velocities for O~VI, as observed, if the line arises
primarily from this location rather than from the entire disk.  The lack
of hydrogen absorption from phases 0.65 to 0.95 might occur if the region
around this hot spot is highly ionized. 

\acknowledgments

APC and PCS gratefully acknowledge support through NASA grant NAG5-8805.

\clearpage

\begin{deluxetable}{lccr}
\tablecaption{FUSE Observations of RX~J0019.8+2156}
\tablehead{
\colhead{Orbit} &\colhead{Photometric phase$^a$} &\colhead{HJD (mid)}
&\colhead{Exposure }\\
&&\colhead{2451700+} &(sec) }
\startdata
1 &0.475 &52.922 &2481\nl
2 &0.584 &52.993 &2191\nl
3 &0.694 &53.066 &1871\nl
4 &0.799 &53.135 &1590\nl
5 &0.907 &53.206 &1310\nl
6 &0.966 &53.246 &418\nl
7 &0.013 &53.277 &1160\nl
8 &0.075 &53.318 &850\nl
9 &0.120 &53.347 &920\nl
10 &0.184 &53.389 &1220\nl
11 &0.227 &53.418 &701\nl
12 &0.293 &53.461 &1721\nl
13 &0.327 &53.4784 &1320\nl
14 &0.414 &53.541 &3461\nl
15 &0.519 &53.610 &3461\nl
16 &0.622 &53.679 &3301\nl
\enddata
\tablenotetext{a}{ephemeris: T$_0$=HJD 2449987.8459(16) $+$ 
E$\times$0.6604645(14) days ~ (McGrath et al.\ 2001)}
\end{deluxetable}

\clearpage

\begin{figure}
\caption{(upper two panels) Average FUSE spectrum of QR And with different
amounts of smoothing and airglow removal.  (bottom two panels) For
comparison, the H$_2$ model absorption is shown with two different
smoothings.  } 
\end{figure}

\begin{figure}
\caption{ Light curves for QR And.  The small open circles (upper curve)
are $V$ data obtained on UT 2000 July 28, only a few hours after the FUSE
observations ended.  The small filled circles  are the $V$ data of McGrath
et al.\ (2001).  It is known that this binary shows variations of a few
tenths of a magnitude in its mean brightness level (e.g. Cowley et al.
1998, Deufel et al. 1999).  The lower two curves are the FUSE light curves
extracted from different wavelength regions, as indicated.  Phases are
computed using the ephemeris given by McGrath et al.} 
\end{figure}

\begin{figure}
\caption{ Line profiles of O~VI, 1032\AA, binned through the orbital
phases indicated.  These are compared with line profiles of H$\beta$ and
H$\gamma$ in the same phase bins (from the data of McGrath et al.).  The
FUSE data have been heavily smoothed to spectral resolution $\sim1$\AA;
the optical data had an original resolution of $\sim5$\AA, but many
spectra have been averaged together.  The vertical scales are in units of
the local continuum. } 
\end{figure}

\begin{figure}
\caption{ Velocity and equivalent width measurements from FUSE and optical
spectra binned as in Fig.\ 3.  The measurements are made with respect to
the continuum levels plotted in Fig.\ 3.  In the upper panel, the mean of
H$\gamma$ and H$\beta$ is plotted to simplify the diagram.  The Balmer
velocities are not corrected for the He~II blending, which would be $-$60
km s$^{-1}$ for equal contributions of H and He~II to the emission. } 
\end{figure}

\begin{figure}
\caption{Similar to Fig.\ 3, but showing changes in profiles from the
0.67--0.95 phase bin, when Balmer absorption is not seen.  The phase
0.67--0.95 profiles are shown in the lowest plots.  This figure
illustrates the changing absorption components with binary phase.} 
\end{figure}

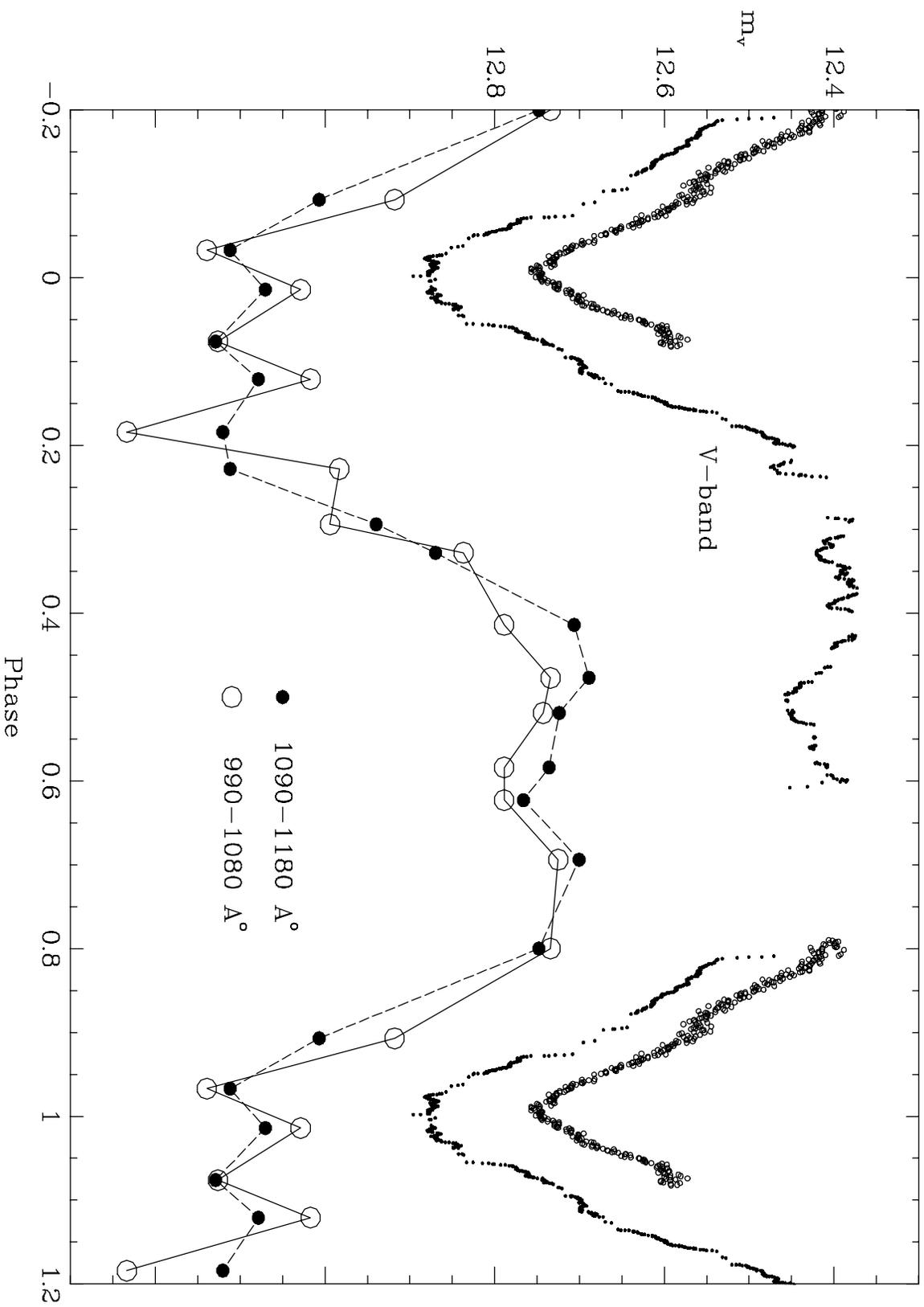
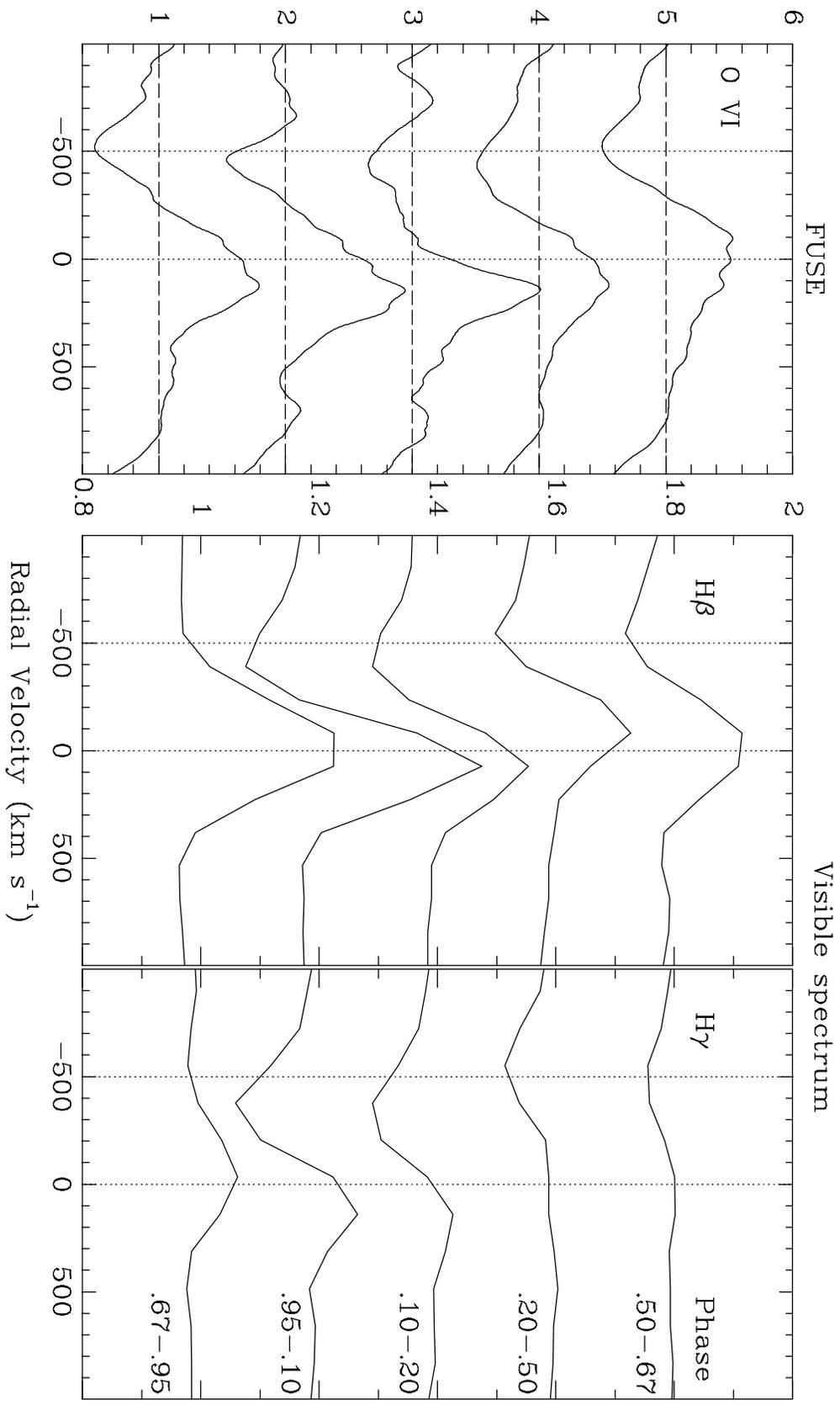
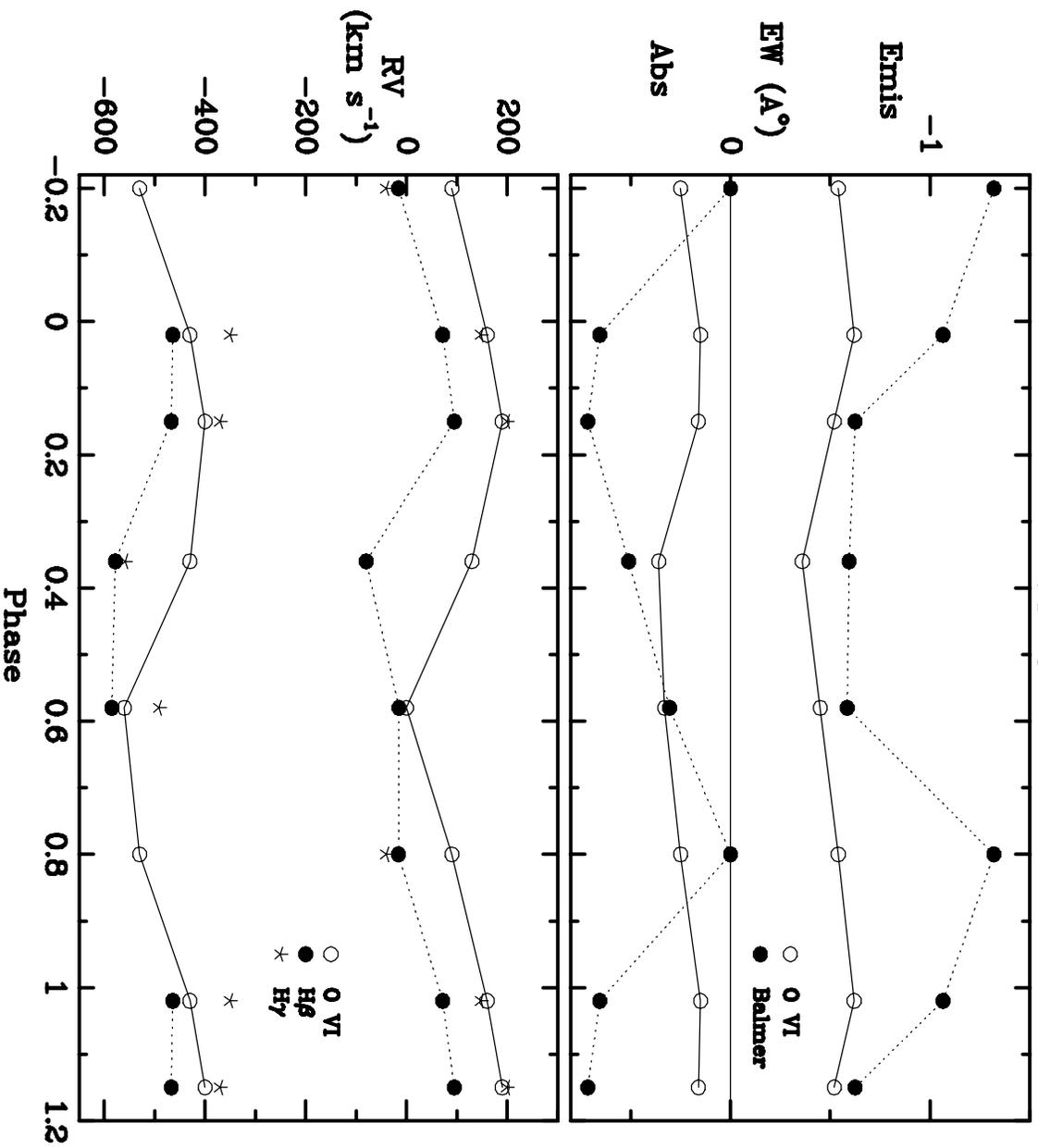
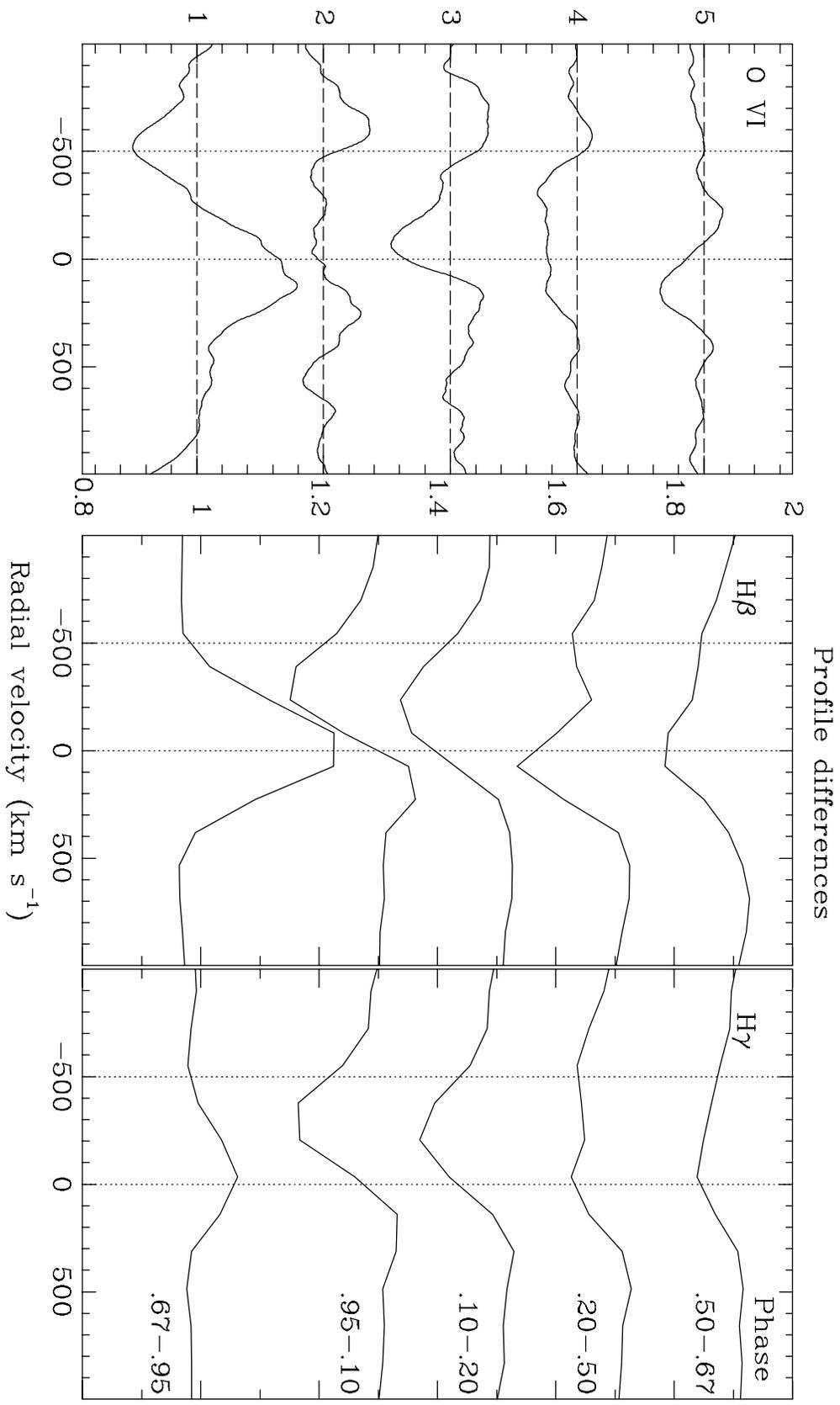
\end{document}